# Implications of the Occurrence of Glitches in Pulsar Free Precession Candidates

D. I. Jones,[1,*] G. Ashton,[1,2] and R. Prix[2]

[1]*Mathematical Sciences, University of Southampton, Southampton SO17 1BJ, United Kingdom*
[2]*Max Planck Institut für Gravitationsphysik (Albert Einstein Institut) and Leibniz Universität Hannover,
30161 Hannover, Germany*
(Received 24 October 2016; published 27 June 2017)

The timing properties of radio pulsars provide a unique probe of neutron star interiors. Recent observations have uncovered quasiperiodicities in the timing and pulse properties of some pulsars, a phenomenon that has often been attributed to free precession of the neutron star, with profound implications for the distribution of superfluidity and superconductivity in the star. We advance this program by developing consistency relations between free precession and pulsars glitches, and we show that there are difficulties in reconciling the two phenomena in some precession candidates. This indicates that the precession model used here needs to be modified or some other phenomenon is at work in producing the quasiperiodicities, or even that there is something missing in terms of our understanding of glitches.

*Introduction.*—Neutron stars are small compact stars, formed at the end point of the life of a sufficiently massive main sequence star. With a mass of $\sim 1.5 M_{\odot}$ contained within a sphere of radius $\sim 12$ km, they represent ready-made laboratories, containing matter at densities well above that of nuclear matter, and are expected to contain superfluids and superconductors.

One way of probing the interiors of neutron stars is through a detailed study of the rotational timing properties of *radio pulsars*. Two sorts of timing features are of particular interest: sudden increases in spin frequency known as *glitches* [1] and *quasiperiodicities* in the observed spin-down rate, correlated in some cases with quasiperiodicities in the pulse beam shape [2].

Pulsar glitches are believed to be caused by sudden changes in the stellar structure. Smaller glitches may be due to *crustquakes*, i.e., fractures in the gradually slowing elastic crust [3], while larger glitches may be related to the sudden release of pinned vorticity in the superfluid interior [4]. The quasiperiodicities may be due to free precession [5,6], which is sensitive to the departure of the star from a spherical configuration. This places highly interesting constraints on the amount of pinned vorticity within the star [6–9]. Also, as shown by Link [10], the long-lived nature of the long period precession also places highly nontrivial constraints on the nature of superconductivity within the star and its location relative to the neutron superfluid.

The stellar characteristics believed to be important for glitches and free precession are intimately connected, and the occurrence of a glitch in a precessing star should allow for a unique test of neutron star theory. Indeed, the likely effect of a crustquake in increasing the precession period was considered by Ruderman [11], while the difficulty of having a large unpinning-type glitch in a precessing star was noted by Link and Cutler [12].

In this Letter, we develop these ideas and apply them to recent pulsar observations. We focus on PSR B1828-11, a particularly well studied free precession candidate [5,6,9, 13–15] that was observed to glitch in 2009 (see Refs. [1,16]), but remark on two other candidates at the end.

Our aim here is to look for consistency (or the lack thereof) of the precession interpretation with the leading models of pulsar glitches, assuming several different models for the stellar deformation and glitch mechanism. This leads to some interesting conclusions, additional to and distinct from those previously obtained from considerations of the precession alone [6,9,10].

*PSR B1828-11.*—We have previously presented a free precession model for PSR B1828-11, with small wobble angle $\theta \approx 3°$ (the angle between the symmetry axis of the biaxial body and its angular momentum) and large magnetic inclination angle $\chi \approx 89°$ (the angle between the symmetry axis and the magnetic axis) [15]. We also carried out a Bayesian model comparison between our free precession model and an alternative model based on magnetospheric switching, and we found that the precession hypothesis was favored. Precessional solutions similar to ours have been found by others [9,14]. In addition, the existence of a second precessional solution of small $\chi$, large $\theta$ was noted by Arzamasskiy *et al.* [14]; we do not analyze that solution here.

More recently, we extended our precession model and obtained the surprising result that the modulation period is steadily decreasing [17]; we will fold this last observation into the considerations of this Letter. The data in our

 P H Y S I C A L   R E V I E W   L E T T E R S 

possession terminate just before the glitch, but we can say that, just before the glitch, the modulation period was $P_{\rm fp} = 468.8$ days $= 1.28$ yr, and we found

$$\epsilon_p \equiv \frac{P}{P_{\rm fp}} \approx 10^{-8}, \qquad \tau_{\rm mod} \equiv \frac{\epsilon_p}{\dot{\epsilon}_p} = 213 \text{ yr}, \qquad (1)$$

where $P$ is the spin period and $\tau_{\rm mod}$ is a time scale characteristic of the decrease in $P_{\rm fp}$.

The glitch itself is reported to have taken place just after the end of our data set, at MJD55041.75 [1]. The fractional frequency change observed at the glitch was $\delta\nu/\nu = 6.2 \times 10^{-9}$, while the fractional frequency derivative change was $\delta\dot{\nu}/\dot{\nu} = 5.2 \times 10^{-3}$ [1]. We will make use of only $\delta\nu/\nu$ in the consistency tests below, as the interpretation of $\delta\dot{\nu}/\dot{\nu}$ is less straightforward, with some glitches (in other pulsars) having values of $\delta\dot{\nu}/\dot{\nu}$ of order unity, or $\delta\dot{\nu}/\dot{\nu} < 0$, neither of which is expected on the basis of the models described below.

*Stellar model.*—Motivated by the long precession periods, we initially do not include a pinned superfluid component. We follow the model described in Jones and Andersson [6] and model an otherwise spherical star of moment of inertia $I_*$ as carrying two quadrupolar perturbations, a "centrifugal" piece, axisymmetric with respect to the (instantaneous) rotation axis, and the other "deformation" piece axisymmetric with respect to some axis fixed within the star. This last perturbation is supported by elastic or magnetic stresses. The moment of inertia tensor can then be shown to be effectively biaxial [6], with principal moments $I_1 = I_2$ and $I_3$, which we will write $I_3 = I_*(1 + \epsilon)$.

In the case of elastic deformations, some insight into the sizes of these deformations can be obtained using a simple energy minimization argument for a steadily rotating star [3,6], which gives $\epsilon$ as the sum of the centrifugal and deformation pieces:

$$\epsilon = \epsilon_\Omega + \epsilon_d \approx \frac{\Omega^2 R^3}{GM} + b\epsilon_{\rm ref}, \qquad (2)$$

for a star of mass $M$ and radius $R$, where $\Omega = 2\pi\nu$, $\epsilon_{\rm ref}$ is the *reference ellipticity* at which the elastic crust would be relaxed, and $b \sim 10^{-7}$, a parameter whose smallness reflects the weakness of elastic forces relative to gravitational ones [18]. The reference shape $\epsilon_{\rm ref}$ will depend upon the "geological" history of the crust and is probably positive, corresponding to the star retaining some memory of its shape when born rotating more rapidly than now. It can be shown that $\Delta I_d \equiv I_3 - I_1 = 3\epsilon_d I_*/2$ [6].

The strain in the star is on the order of $u \sim \epsilon_{\rm ref} - \epsilon$, so that $u \sim \epsilon_{\rm ref}$ for sufficiently slowly spinning stars, and can probably be no larger than $\sim 0.1$ [19].

In the case of magnetic deformations, a simple energy-based estimate leads to

$$\epsilon_d \approx \frac{B^2 R^3}{GM^2/R} \equiv k_{\rm normal} B_{\rm int}^2 \approx 1.9 \times 10^{-12} B_{12}^2 \qquad (3)$$

for nonsuperconducting stars, while

$$\epsilon_d \approx \frac{BH_cR^3}{GM^2/R} \equiv k_{\rm supercon} B_{\rm int} \approx 1.9 \times 10^{-9} B_{12} \qquad (4)$$

for superconducting ones, where $B$ is the internal field strength and $H_c \sim 10^{15}$ G [18].

*Consistency between the glitch and the free precession.*—We will begin by considering a star deformed by some combination of elastic and magnetic strains, with no pinned superfluid component. First, consider the glitch. The star's angular momentum is given by $J = I_*(1 + \epsilon)\Omega$. Angular momentum conservation over the glitch demands $\delta J = 0$, so we have

$$(1 + \epsilon)\delta\nu + \nu\delta\epsilon \Rightarrow \delta\epsilon \approx -\frac{\delta\nu}{\nu}. \qquad (5)$$

We can break up the total ellipticity as given in Eq. (2), so that $\delta\epsilon = \delta\epsilon_\Omega + \delta\epsilon_d$. Given that $\epsilon_\Omega \sim \nu^2$, we have $\delta\epsilon_\Omega \approx 2(\delta\nu/\nu)\epsilon_\Omega$. This term is negligible, leaving $\delta\epsilon \approx \delta\epsilon_d$, and we finally have

$$\delta\epsilon_d \approx -\frac{\delta\nu}{\nu}. \qquad (6)$$

Now turn to the free precession. Rigid body dynamics then says that, for small $\theta$,

$$\epsilon_p \equiv \frac{P}{P_{\rm fp}} = \frac{|\Delta I_d|}{I_{\rm prec}} = \frac{|\Delta I_d|}{I_*}\frac{I_*}{I_{\rm prec}} = \frac{3}{2}|\epsilon_d|\frac{I_*}{I_{\rm prec}}, \qquad (7)$$

where $I_{\rm prec}$ is the portion of the spherical part of the moment of inertia that participates in the free precession [6]. The size of $I_{\rm prec}$ depends upon how tightly the interior fluid is coupled to the crust. We can certainly expect $I_{\rm crust} < I_{\rm prec} < I_*$ [18]. Note that we have taken the modulus of $\epsilon_p$ and $\Delta I_d$, as an observation of the free precession period alone cannot distinguish between the oblate and prolate cases [15]. Rearranging leads to

$$|\epsilon_d| = \frac{2}{3}\frac{P}{P_{\rm fp}}\frac{I_{\rm prec}}{I_*} = 6.67 \times 10^{-9}\left(\frac{P/P_{\rm fp}}{10^{-8}}\right)\frac{I_{\rm prec}}{I_*}. \qquad (8)$$

By a simple addition, we can then calculate the deformation *after* the glitch:

$$\epsilon_{d,{\rm after}} = \epsilon_{d,{\rm before}} + \delta\epsilon_d. \qquad (9)$$

Given that the spin-up glitch occurred such that $\delta\epsilon_d < 0$, we will assume that the deformation is oblate ($\epsilon_d > 0$), not prolate ($\epsilon_d < 0$); i.e., the glitch represented a decrease in the magnitude of the deformation, breaking the degeneracy inherent in the interpretation of the precession. For pure elastic or pure magnetic deformations, this seems natural; if the deformation is sourced by a combination of elastic and magnetic strains, this assumption is less safe, as a (dominantly) toroidal field would produce a prolate deformation [20].

We can use the formulas given above to obtain

$$\epsilon_{d,\text{after}} = \frac{2}{3} \frac{P}{P_{\text{fp,before}}} \frac{I_{\text{prec}}}{I_*} - \frac{\delta\nu}{\nu}. \quad (10)$$

In line with our assumption of oblateness described above, we will require $\epsilon_{d,\text{after}} > 0$; i.e., the glitch can relieve no more strain than was originally present in the star. This leads to a *lower bound* on $I_{\text{prec}}/I_*$, such that

$$\frac{3}{2} \frac{\delta\nu/\nu}{P/P_{\text{fp}}} \le \frac{I_{\text{prec}}}{I_*} \le 1 \Rightarrow 0.93 \le \frac{I_{\text{prec}}}{I_*} \le 1; \quad (11)$$

i.e., at least 93% of the total moment of inertia must participate in the precession. This can be regarded as a consistency test: a lower bound on $I_{\text{prec}}/I_*$ in excess of unity would point to a lack of consistency between the glitch and the precession, assuming that pinned superfluidity plays no role in either.

Combining this with Eq. (8), we obtain constraints on $\epsilon_d$ just before the glitch:

$$\frac{\delta\nu}{\nu} \le \epsilon_{d,\text{before}} \le \frac{2}{3} \frac{P}{P_{\text{fp,before}}}, \quad (12)$$

$$\Rightarrow 6.2 \times 10^{-9} \le \epsilon_{d,\text{before}} \le 6.67 \times 10^{-9}, \quad (13)$$

an impressively tight range.

We can also constrain the range of $\epsilon_d$ and $P_{\text{fp}}$ after the glitch. It can be shown, using the results above, that

$$0 \le \epsilon_{d,\text{after}} \le \frac{2}{3} \frac{P}{P_{\text{fp,before}}} - \frac{\delta\nu}{\nu}, \quad (14)$$

$$\Rightarrow 14.3 \le \frac{P_{\text{fp,after}}}{P_{\text{fp,before}}} = \frac{\epsilon_{d,\text{before}}}{\epsilon_{d,\text{after}}} \le \infty; \quad (15)$$

i.e., the deformation must be reduced be a factor of at least ∼14 after the glitch. This corresponds to a postglitch precession period $P_{\text{fp,after}} > 18.4$ yr.

We are not able to test this last prediction directly, as our data set stopped just prior to the glitch. Some relevant timing data is in fact given in Brook *et al.* [21] and Kerr *et al.* [22], where a small amount of postglitch data are presented. Visual inspection makes clear that the quasiperiodicity was *not* significantly affected by the glitch, but a more careful analysis is needed to quantitatively estimate (or bound) any change in modulation period.

From Eq. (7), we can interpret the increasing value of $\epsilon_p$ in terms of an increasing deformation $\epsilon_d$. The gradually increasing deformation $\epsilon_d$ would replenish the deformation undone at the glitch, $\delta\epsilon_d$, in a time scale $\Delta t_{\text{replenish}} = |\delta\epsilon_d|/|\dot{\epsilon}_d|$. Using Eqs. (6) and (7), we obtain

$$\Delta t_{\text{replenish}} = \frac{\delta\nu}{\nu} \frac{3}{2} \frac{I_*}{I_{\text{prec}}} \frac{1}{\dot{\epsilon}_p} = 198 \text{ yr} \frac{I_*}{I_{\text{prec}}}. \quad (16)$$

One can hypothesize that some (unknown) mechanism produces a gradually increasing deformation between glitches, with the deformation being reset to smaller values in periodic glitches.

*Specializing to elastic deformations.*—Using the relation $\epsilon_d = b\epsilon_{\text{ref}}$ and Eq. (12), we can make a statement concerning the preglitch reference shape:

$$\frac{1}{b} \frac{\delta\nu}{\nu} \le \epsilon_{\text{ref,before}} \le \frac{1}{b} \frac{2}{3} \frac{P}{P_{\text{fp,before}}}, \quad (17)$$

$$\Rightarrow \frac{6.2 \times 10^{-2}}{b_{-7}} \le \epsilon_{\text{ref,before}} \le \frac{6.67 \times 10^{-2}}{b_{-7}}. \quad (18)$$

These values imply a similarly high value for the strain in the crust, right at the upper end of the values obtained by Horowitz and Kadau [19], possibly suggesting that it is indeed crustal fracture that triggers the glitch. However, the gradual decrease in modulation period implies a gradual increase in the reference shape and in the corresponding elastic strains. This is difficult to understand, as plastic flow processes can be expected to always tend to decrease the strain, not increase it.

*Specializing to magnetic deformations.*—Now suppose that the deformation is produced by a superconducting core. Combining Eqs. (4) and (12), we can constrain the *internal* magnetic field strength prior to the glitch:

$$\frac{1}{k_{\text{supercon}}} \frac{\delta\nu}{\nu} \le B_{\text{int,before}} \le \frac{1}{k_{\text{supercon}}} \frac{2}{3} \frac{P}{P_{\text{fp,before}}}, \quad (19)$$

which (in units of $10^{12}$ G) evaluates to

$$3.26 \le B^{\text{int}}_{12,\text{before}} \le 3.51. \quad (20)$$

[Equation (4) is a rough estimate, so this inequality is accurate only to an overall multiplicative factor of order unity.] We can compare this with the value of the external dipole field, $B_{12,\text{external}} \approx 5.0$, inferred from the spin-down rate (see Refs. [23,24]). In contrast with the elastic strains required to explain the precession, these estimated internal magnetic field strengths are sensible and close to the inferred external field strength.

Given that we are now assuming that it is magnetic strains alone that deform the star, we are compelled to explore the unconventional idea that the glitch represents a sudden decrease in the star's internal magnetic field. Then $\delta\epsilon_d = k_{\text{supercon}}\delta B_{\text{int}}$, so that Eq. (14) gives

$$0 \le B_{\text{int,after}} \le \frac{1}{k_{\text{supercon}}} \left( \frac{2}{3} \frac{P}{P_{\text{fp}}} - \frac{\delta\nu}{\nu} \right), \quad (21)$$

$$\Rightarrow 0 \le B^{\text{int}}_{12,\text{after}} \le 0.246. \quad (22)$$

This is problematic: the large reduction of deformation at the glitch requires a large reduction in the interior magnetic field strength. This seems rather unlikely, particularly given that there has been no corresponding large reduction in the (inferred) external field strength following the glitch.

The decreasing $P_{\text{fp}}$ reported above can be interpreted as a gradually increasing internal magnetic field (as distinct from the sudden decrease $\delta B_{\text{int}}$ in the field that might accompany the glitch), growing on a time scale $\sim 213$ yr. That the field should be increasing—and changing on such a short time scale—is theoretically unexpected (see, e.g., Pons and Geppert [25]). One can also see that there is no accompanying rapid increase in the external field. Assuming a power law spin-down $\dot{\nu} \propto B_{\text{ext}}^2 \nu^n$, the braking index $n_{\text{obs}} \equiv \nu \ddot{\nu}/\dot{\nu}^2$ is given by $n_{\text{obs}} = n + 2\tau_{\text{sd}}/\tau_{B-\text{ext}} \approx n + 1005$, much greater than the actual value $n_{\text{obs}} \approx 16$ [15], showing that the rapid field increase needs to be confined to the interior.

In the case of a normal interior, one can carry through a nearly identical analysis using Eq. (3) in place of Eq. (4), obtaining broadly similar results, with slightly higher inferred magnetic fields: $57.1 \leq B_{12,\text{before}}^{\text{int}} \leq 59.2$, and $0 \leq B_{12,\text{after}}^{\text{int}} \leq 15.7$.

*Allowing for pinned superfluidity.*—Given the difficulties encountered above, let us turn to a model based on superfluid pinning. If a star contains a perfectly pinned superfluid component with moment of inertia $I_{\text{PSF}}$, then Shaham [7] showed that (neglecting the contribution from elastic or magnetic deformations considered above)

$$\frac{P}{P_{\text{fp}}} = \frac{I_{\text{PSF}}}{I_{\text{prec}}} = \frac{I_{\text{PSF}}}{I_*} \frac{I_*}{I_{\text{prec}}}. \qquad (23)$$

Using the observed preglitch modulation period of PSR B1828-11,

$$\frac{I_{\text{PSF,before}}}{I_*} = \frac{I_{\text{prec}}}{I_*} \frac{P}{P_{\text{fp,before}}} < 10^{-8}. \qquad (24)$$

This is the well-known result that the free precession of PSR B1828-11 places a tight constraint on the amount of pinned superfluid in the star [6,9]. This is to be compared with the expectation that, in fact, $I_{\text{PSF}}/I_* \sim 10^{-2}$, which comes both from the superfluid model of glitches and from the theoretical expectation of how much superfluid coexists with the inner crust; see, e.g., Ref. [26].

A way out of this problem was proposed by Link and Cutler [12], who argued that the precession motion itself could cause superfluid vortices to unpin from the crust, aside from a small region near the rotational equator. The following rather pretty picture then suggests itself: at some time prior to observations, PSR B1828-11 underwent a glitch that set the star into precession, braking all or most of the superfluid pinning in the process. The gradually decreasing precession period could then be interpreted as a gradual reestablishment of the pinning.

However, a little analysis reveals difficulties with this interpretation. If we label the moments of inertia of the pinned superfluid before (after) the glitch as $I_{\text{PSF,before}}$ ($I_{\text{PSF,after}}$), angular momentum conservation at the glitch gives

$$0 \approx \delta I_{\text{unpin}} \delta\nu_{\text{PSF}} + I_* \delta\nu, \qquad (25)$$

where $\delta\nu$ is the observed spin change, $\delta\nu_{\text{PSF}}$ is the spin change experienced by the (unseen) portion of superfluid that unpins, and $\delta I_{\text{unpin}} = I_{\text{PSF,before}} - I_{\text{PSF,after}}$.

Rearranging and rewriting slightly leads to

$$-\delta\nu_{\text{PSF}} = \frac{I_*}{\delta I_{\text{unpin}}}\delta\nu = \frac{I_*}{I_{\text{PSF,before}}} \frac{I_{\text{PSF,before}}}{\delta I_{\text{unpin}}}\delta\nu. \qquad (26)$$

Using Eq. (24) to eliminate $I_{\text{PSF,before}}$ in favor of $P_{\text{fp,before}}$, we get

$$-\delta\nu_{\text{PSF}} = \frac{P_{\text{fp}}}{P} \frac{I_*}{I_{\text{prec}}} \frac{I_{\text{PSF,before}}}{\delta I_{\text{unpin}}}\delta\nu. \qquad (27)$$

Given that $I_*/I_{\text{prec}} > 1$ and $I_{\text{PSF,before}}/\delta I_{\text{unpin}} > 1$, this equation immediately leads to a lower bound on $-\delta\nu_{\text{PSF}}$. Note also that the lag between the pinned superfluid and the rest of the star is $\nu_{\text{lag}} = \nu_{\text{PSF}} - \nu$, so the change in lag at a glitch is $\delta\nu_{\text{lag}} = \delta\nu_{\text{PSF}} - \delta\nu \approx \delta\nu_{\text{PSF}}$. It follows that our lower bound on the change in rotation rate of the pinned superfluid above is also a lower bound on the change in lag between the pinned superfluid and the rest of the star, so there is, therefore, also a lower bound on the actual preglitch lag between the two stellar components:

$$\nu_{\text{lag}} > \frac{P_{\text{fp}}}{P}\delta\nu = 1.53 \text{ Hz.} \qquad (28)$$

This is problematically large. Application of Eq. (25) to, for instance, the Vela pulsar, with $\nu/\dot{\nu} \sim 10^{-6}$ and $\delta I_{\text{unpin}}/I_* \sim 10^{-2}$, leads to $\nu_{\text{lag}} \sim 10^{-3}$ Hz, 3 orders of magnitude smaller than for PSR B1828-11. More problematically, the lower bound on the lag for PSR B1828-11 is only slightly less than its current spin frequency ($\nu = 2.47$ Hz), so the star would had to have spun down without glitching for most of its lifetime to accumulate such a lag.

As was the case for elastic or magnetic deformations, there will, in the pinned superfluid case, be an increase in the free precession period coincident with the glitch, in this case by a factor $I_{\text{PSF}}/(I_{\text{PSF}} - \delta I_{\text{unpin}})$, but we cannot quantity how large this increase will be, as we cannot constrain $\delta I_{\text{unpin}}$, only the product $\delta I_{\text{unpin}}\delta\nu_{\text{PSF}}$; see Eq. (25).

We can additionally note that the time scale $\Delta t_{\text{repin}}$ for the gradual repinning to reestablish a reservoir of pinned superfluid of moment of inertia $\Delta t_{\text{repin}}$ is long. From Eq. (23), we have $\dot{I}_{\text{PSF}} = I_{\text{prec}}\dot{\epsilon}_p$, so

$$\Delta t_{\text{repin}} = \frac{\Delta I_{\text{repin}}}{\dot{I}_{\text{PSF}}} = 2.13 \times 10^8 \text{ yr} \frac{\Delta I_{\text{repin}}/I_*}{10^{-2}} \frac{I_*}{I_{\text{prec}}}, \qquad (29)$$

implying that such unpinning events have to be rare, as PSR B1828-11 will not build up a typically sized pinned superfluid reservoir for a long time to come.

*Other precession candidates.*—We will comment briefly on two other free precession candidates. PSR B0919+06 displays correlated quasiperiodicities in spin-down and beamwidth and has also glitched [27]. Using data from Perera *et al.* [27] and assuming no pinned superfluidity, Eq. (11) gives the (nonsensical) result $I_{prec}/I_* > 213$. When we assume pinned superfluidity to be relevant, Eq. (28) gives a (huge) lower bound on the preglitch lag of $\nu_{lag} > 327$ Hz. PSR J1646-4346 was identified as a precession candidate by Kerr *et al.* [22], who also reported a glitch. Using their data, we have $I_{prec}/I_* > 572$, or $\nu_{lag} > 1660$ Hz. In both cases, there is a lack of consistency, with the glitch too large to be accommodated within our precession model.

*Summary and discussion.*—There are significant problems in reconciling the free precession interpretation of the quasiperiodicities in PSR B1828-11 with the glitch that occurred in this pulsar. Depending upon the model assumed, the problems lie in the postglitch precession period apparently not increasing, the inferred elastic strains being too large and increasing, the internal magnetic field having to change rapidly with no corresponding evolution in the external field, or the inferred lag between the crust and pinned superfluid being too large. There are even greater consistency issues in at least two other (albeit less well studied) precession candidates.

On the basis of these considerations, it would seem that the particular free precession model used here (small $\theta$ with large $\chi, \epsilon_d > 0$) is not consistent with the observed glitches. What can one conclude? One possibility is that precession is nevertheless the mechanism responsible for producing the modulations, but the particular realization of precession used here is not the correct one. It would be interesting to explore the large $\theta$, small $\chi$ precession solution described in Arzamasskiy *et al.* [14], although this would inevitably involve modeling the star as triaxial. Similarly, prolate scenarios, perhaps with both crustal strain and magnetic fields playing a role, might be relevant. It may be of interest to relax our assumption of perfect pinning by allowing for vortex creep, as considered in Ref. [28], but the coupling between the vortices and crust would presumably have to be very weak to recover the long precession period, and yet be strong enough to build up sufficient lag to trigger the glitch. Finally, it may be that this lack of consistency is evidence of flaws in our understanding of glitches.

Alternatively, the magnetospheric switching or planetary companion(s) might be needed. The latter is attractive in that it not only provides a clock mechanism, but the slowly decreasing modulation period of PSR B1828-11 might have a natural explanation in the gradual decay of the orbit.

In the immediate term, the most useful task would be to perform an analysis of the timing data for glitching precession candidates, with a view to setting upper limits on changes in the quasiperiodic behavior coincident with the glitch. Ideally, this would be done also allowing for the secular variation in modulation period described here for PSR B1828-11. As we have argued here, such changes provide a potentially powerful tool for probing the dynamics and structure of neutron stars.

We thank Nils Andersson and Kostas Glampedakis for their comments on this manuscript. D. I. J. acknowledges support from Science and Technologies Funding Council via Grant No. ST/H002359/1, and also travel support from NewCompStar (a COST-funded research networking program).

*d.i.jones@soton.ac.uk